\documentstyle[amsmath,epsfig,wrapfig]{aipproc}
\begin{document}

\pagestyle{empty}

\begin{flushleft}
\Large
{SAGA-HE-161-00
\hfill December 11, 2000}  \\
\end{flushleft}
 
\vspace{2.6cm}
 
\begin{center}
 
\huge{{\bf Polarized structure functions}} \\
\vspace{0.2cm}

\huge{{\bf of the deuteron}} \\

\vspace{1.5cm}
 
\LARGE
{\ S. Kumano $^*$} \\
 
\vspace{0.3cm}
  
\LARGE
{Department of Physics}         \\
 
\LARGE
{Saga University}      \\
 
\LARGE
{Saga 840-8502, Japan} \\

\vspace{2.0cm}
 
\LARGE
{Talk given at the 14th International} \\

\vspace{0.1cm}

{Spin Physics Symposium} \\

\vspace{0.35cm}

\Large
{Osaka, Japan, October 16 -- 21, 2000} \\

\vspace{0.05cm}

{(talk on Oct. 16, 2000) }  \\
 
\end{center}
 
\vspace{0.7cm}

\vfill
 
\noindent
{\rule{6.0cm}{0.1mm}} \\
 
\vspace{-0.3cm}
\normalsize
\noindent
{* Email: kumanos@cc.saga-u.ac.jp.
          Information on his research is available at} \\

\vspace{-0.44cm}
\noindent
{\ \ \ http://www-hs.phys.saga-u.ac.jp.}  \\

\vspace{+0.1cm}
\hfill
{\large to be published in AIP proceedings}

\vfill\eject

\normalsize

\title{Polarized structure functions \\ of the deuteron}

\author{S. Kumano}
\address{Department of Physics, Saga University, Saga 840-8502, Japan}

\maketitle

\begin{abstract}
Physics of spin-1 hadron is an unexplored topic in the high-energy region
although spin-1/2 physics has been well investigated in the last decade.
It is important to test our knowledge of hadron spin structure in 
a quite different field of spin physics.
We discuss tensor structure functions, which do not exist for
the spin-1/2 nucleon, in lepton scattering and in hadron reactions
such as the polarized proton-deuteron Drell-Yan process.
\end{abstract}

\section*{Introduction}

High-energy spin physics is entering into a new era in the sense that
the difference between a naive quark-model prediction and
experimental data is now roughly understood and that much detailed studies
are in progress with the completion of the RHIC facility.
So far, spin-structure studies have been focused on the spin-1/2
nucleon. For higher-spin hadrons, we know that different
spin physics exists such as the tensor structure in the deuteron.
In order to test our understanding of the hadron structure, it is 
important to investigate other spin quantities. In this sense,
spin-1 hadrons, for example the deuteron, should be suitable
for future investigations. 

In this paper, the present status is explained for the deuteron
spin structure in the high-energy region. First, polarized
electron-deuteron scattering is discussed briefly in connection
with the deuteron spin physics. In particular, new polarized
structure functions are introduced.
Then, as an alternative method, the polarized proton-deuteron
Drell-Yan process is discussed for finding the new distributions.

\section*{Spin-1 structure in electron scattering}

A general formalism of the deep inelastic electron-deuteron scattering
is discussed in Ref. \cite{mit-b1}. It suggests that there exist
additional structure functions in comparison with the electron-proton
scattering due to the spin-1 nature of the deuteron. 
We find that there are eight independent
amplitudes for $ \gamma (h_1)+d (H_1) \rightarrow \gamma (h_2)+d (H_2)$,
where $h_{1,\ 2}$ and $H_{1,\ 2}$ are helicities, 
by using momentum conservation, parity invariance,
and time-reversal invariance. 
Four of them are usual structure functions: $F_1$, $F_2$, $g_1$,
and $g_2$, which exist in the spin-1/2 proton. In addition, there
are four new structure functions: $b_1$, $b_2$, $b_3$, and $b_4$.
The $b_3$ and $b_4$ are higher-twist functions and $b_2$ is related
to $b_1$ by the ``Callan-Gross type" relation $b_2 = 2x b_1$
in the Bjorken scaling limit, so that the most essential part
of new physics is contained in $b_1$. A phenomenological sum rule
$\int dx \, b_1 (x) = 0$
was proposed \cite{ck} if there is no tensor polarization for antiquark
distributions; however, it does not mean that $b_1(x)$ itself
vanishes. This sum rule is similar to the Gottfried sum rule
in a way \cite{skpr}. Because there is no solid reason why
the antiquark tensor polarization vanishes, the violation of
the above sum rule could suggest a finite tensor polarization
as the Gottfried-sum-rule violation indicated a finite $\bar u/\bar d$
asymmetry. 

There are three twist-2 structure functions:
$F_1$, $g_1$, and $b_1$, and they are related to electron-deuteron
scattering cross sections as
\begin{align}
& F_1 \propto  \langle d \sigma \rangle \ ,
\nonumber \\
& g_1 \propto  d \sigma (\uparrow,+1) - d \sigma (\uparrow,-1)  \ ,
\nonumber \\
& b_1 \propto  d \sigma (0) - [ \, d \sigma (+1) + d \sigma (-1) \, ]/2  \ .
\end{align}
Here, $d \sigma (\uparrow, +1)$ and $d \sigma (\uparrow, -1)$
indicate spin parallel and anti-parallel cross sections, respectively.
On the other hand, $\sigma (0)$ and $d \sigma (\pm 1)$ indicate
cross sections with the deuteron spin state $0$ and $\pm 1$.
This fact means that the electron does not have to be polarized
for measuring $b_1$. 
There are a number of predictions on the $x$ dependence of $b_1$, but
we cannot discuss the model results because of the space limitation.
The interested reader may find a list of papers in the reference
section of Ref.\cite{hk}. At this stage, $b_1$ is expected to be
quite small, typically $b_1/F_1\sim$ a few percent or much less.
However, our experience of $g_1$ implies that the predictions
could be wrong. In particular, the tensor structure in the high-energy
region is still an unexamined topic experimentally.  
The HERMES experiment for measuring $b_1$ is 
in progress and its results will be reported in the near future. 
Therefore, the $b_1$ physics could become one of popular
topics in hadron physics.

\section*{Proton-deuteron Drell-Yan process}

In addition to the lepton scattering studies, it is, in principle, possible
to investigate the same spin physics in hadron facilities. 
However, there are only a few theoretical studies on the spin-1 physics
in hadron collisions. So far, the polarized proton-deuteron ($pd$) Drell-Yan
reaction has been investigated in connection with the tensor structure
of the deuteron \cite{hk,km}.

Imposing the conditions of Hermiticity, parity conservation, and
time-reversal invariance, we found many structure functions
in the $pd$ Drell-Yan. Even after integrating the cross section
over the transverse momentum of the virtual photon $\vec Q_T$ or
after taking the limit $Q_T \rightarrow 0$, there are 22 functions \cite{hk}:
\vfill\eject
\begin{alignat}{2}
& {\rm unpolarized:}      & & W_{0,0}, \         W_{2,0} \
\nonumber \\
& {\rm L/T \ polarized:}    & & V_{0,0}^{LL}, \    V_{0,0}^{TT}, \ 
                        V_{2,0}^{LL}, \    V_{2,0}^{TT}, \
                        U_{2,1}^{T U}, \   U_{2,1}^{U T}, \
                        U_{2,1}^{TL},  \   U_{2,1}^{LT}, \
                        U_{2,2}^{TT}  \   
\nonumber \\
& {\rm tensor \ polarized:} \ & & V_{0,0}^{U Q_0}, \ V_{0,0}^{TQ_1}, \ 
                        V_{2,0}^{U Q_0}, \ V_{2,0}^{TQ_1}, \
                        U_{2,1}^{TQ_0},  \ U_{2,1}^{LQ_1}, \
                        U_{2,1}^{TQ_2},  \ U_{2,1}^{U Q_1}, 
\nonumber \\
& \ & &                 U_{2,2}^{U Q_2}, \ U_{2,2}^{T Q_1}, \
                        U_{2,2}^{LQ_2}.
\label{pd-sf}
\end{alignat}
The subscripts of $F_{L,M}^{pol_p \, pol_d}$ ($F$=$W$, $V$, or $U$)
indicate that it is obtained by the integration
$\int d\Omega \,  Y_{LM} \, d\sigma/(d^4Q \,d\Omega)$.
The superscripts indicate the polarization states of the proton
and deuteron. The notations $U$, $L$, and $T$ indicate unpolarized,
longitudinally polarized, and transversely polarized states.
The quadrupole polarizations $Q_0$, $Q_1$, and $Q_2$ are specific
in the reactions with a spin-1 hadron, and
they are associated with the quadrupole terms:
$Q_0$ for the term  $3 \, cos^2 \beta_B -1 \sim Y_{20}$,
$Q_1$ for $sin \beta_B \, cos \beta_B \sim Y_{21}$, and
$Q_2$ for $sin^2 \beta_B \sim Y_{22}$ with the polarization
angle of the deuteron $\beta_B$.
The structure functions in the categories ``unpolarized" and 
``L/T polarized" exist in the proton-proton ($pp$) Drell-Yan.
The new functions are listed as ``tensor polarized" and they
are associated with the spin-1 deuteron.

In defining the spin asymmetry, we should be careful about the
tensor contributions. If it is defined in the usual way, for example
$A_{LL}^{usual}=[ \sigma(\uparrow , -1) -
           \sigma(\uparrow , $ $  +1) ] /
        [ \sigma(\uparrow , -1) + \sigma(\uparrow , +1) ]$,
the denominator is not equal to the unpolarized cross section because of
\begin{equation}
2 < \! \sigma \! > \, =  \sigma(\uparrow , +1) + \sigma(\uparrow , -1) 
   + \frac{1}{3} \,  [ \, 2 \, \sigma(\uparrow , 0)
        - \sigma(\uparrow , +1) - \sigma(\uparrow , -1) \, ]    \ .
\end{equation}
In order to avoid the above tensor contribution, the unpolarized cross section
is used for the denominator in Ref.\cite{hk} as a ``theoretical" definition
of the spin asymmetry. Using the obtained $pd$ cross section with
the structure functions in Eq.(\ref{pd-sf}), we find the following
spin asymmetries:
\begin{alignat}{8}
& < \! \sigma \! >, \ \ & & 
A_{LL}, \ \             & &
A_{TT}, \ \             & &
A_{LT}, \ \             & &
A_{TL}, \ \             & &
A_{UT}, \ \             & &
A_{TU}, \ \             & &
  \                \nonumber \\
& A_{UQ_0}, \ \         & &     
A_{TQ_0}, \ \           & &
A_{UQ_1}, \ \           & &
A_{LQ_1}, \ \           & &
A_{TQ_1}, \ \           & &
A_{UQ_2}, \ \           & &
A_{LQ_2}, \ \           & &
A_{TQ_2}.
\end{alignat}
Those in the first line exist in the $pp$ reaction. However, the ones
in the second line are specific for the $pd$ Drell-Yan, and they
are related to the tensor polarizations $Q_0$, $Q_1$, and $Q_2$.
For example, $A_{UQ_0}$, $A_{UQ_1}$, and $A_{UQ_2}$ are given as
\small
\begin{align}
A_{UQ_0} & = \frac{1}{2 < \! \sigma \! >} \,        
         \bigg [ \, \sigma(\bullet , 0_L)
            - \frac{ \sigma(\bullet , +1_L) 
                    +\sigma(\bullet , -1_L) }{2} \, \bigg ]   
               =  \frac{ 2 \, V_{0,0}^{UQ_0} 
                          + (\frac{1}{3}-cos^2 \theta ) \, 
                            V_{2,0}^{UQ_0} }
                 {  2 \, W_{0,0}
                          + (\frac{1}{3}-cos^2 \theta ) \, W_{2,0}  }  \ ,
\nonumber \\
A_{UQ_1} & = \frac{ \sigma(\bullet , I_1) - \sigma(\bullet , I_3) }
                    { 2 < \! \sigma \! > }
              =  \frac{ sin\theta \, cos\theta \, 
                             sin \phi \, U_{2,1}^{UQ_1} }
                 { 2 \, W_{0,0} 
                          + (\frac{1}{3}-cos^2 \theta ) \, W_{2,0} }  \ ,
\nonumber \\
A_{UQ_2} & = \frac{1}{2 < \! \sigma \! >} \,        
         \bigg [ \, 
             \frac{ \sigma(\bullet , \phi_B=0) 
                         + \sigma(\bullet , \phi_B=\pi) }{2}
            - \frac{ \sigma(\bullet , \phi_B=\pi/2) 
                   + \sigma(\bullet , \phi_B=3\pi/2) }{2} \, \bigg ]   
\nonumber \\
              & =  \frac{  sin^2 \theta \, cos 2 \phi \,
                           U_{2,2}^{U Q_2} }
                 { 2 \, W_{0,0} 
                          +  \, (\frac{1}{3}-cos^2 \theta ) \, W_{2,0} }
\ . 
\end{align}
\normalsize 
Here, $\theta$ and $\phi$ are polar and azimuthal angles of the 
${\ell^+}$ momentum.
The solid circle indicates that the proton is unpolarized.
The three tensor polarizations are illustrated in Fig.1.

\vspace{-0.7cm}
\noindent
\begin{figure}[h]
\parbox[t]{0.31\textwidth}{
   \begin{center}
       \epsfig{file=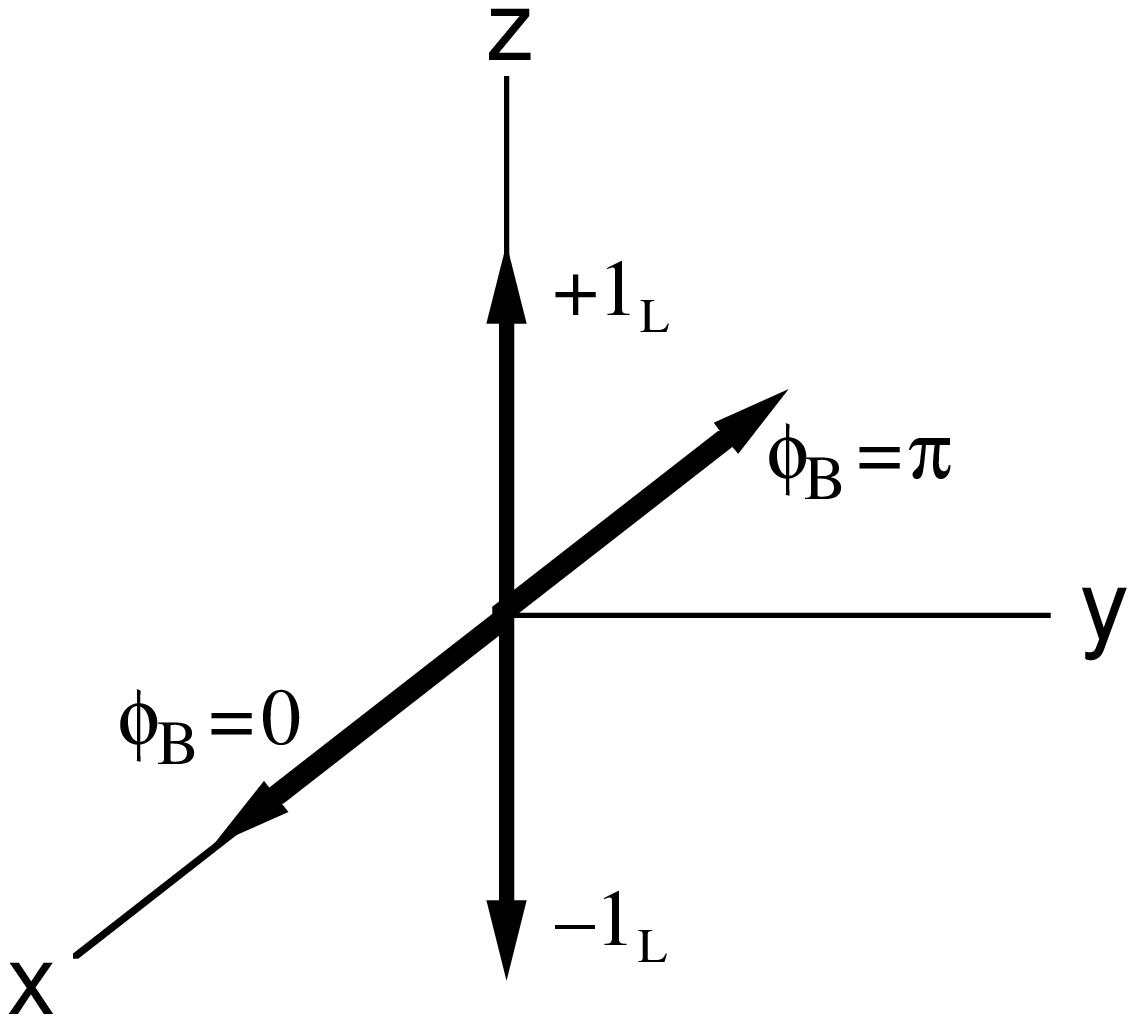,width=4.5cm}
   \end{center}
}\hfill
\parbox[t]{0.31\textwidth}{
   \begin{center}
       \epsfig{file=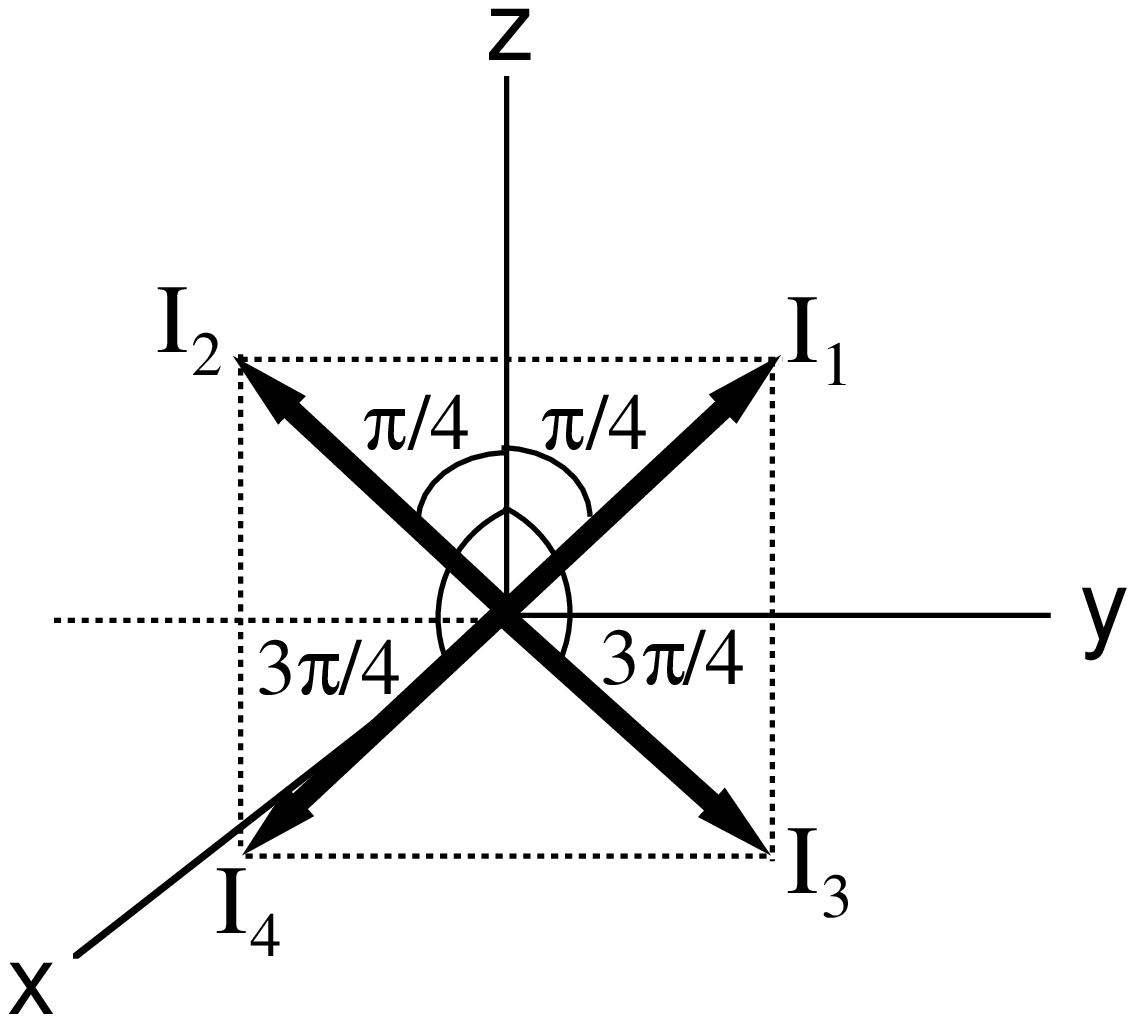,width=4.5cm}
   \end{center}
}\hfill
\parbox[t]{0.31\textwidth}{
   \begin{center}
       \epsfig{file=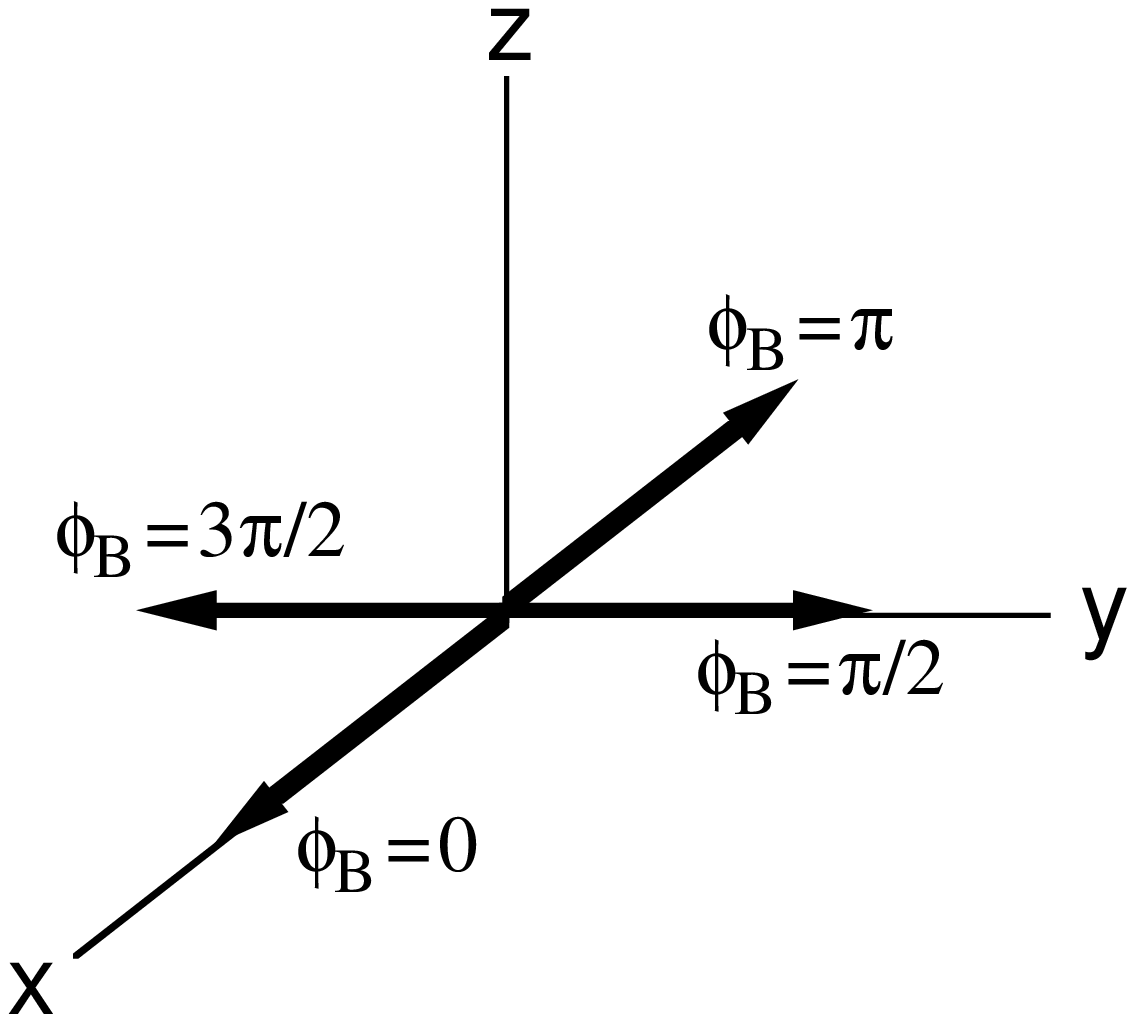,width=4.5cm}
   \end{center}
}
\vspace{-0.1cm}
\caption{$Q_0$, $Q_1$, and $Q_2$ polarizations [4].}
\label{fig:q012}
\end{figure}
\vspace{-0.2cm}

\noindent
The $Q_0$ polarization asymmetry is in principle the same as the one
in defining the tensor structure function $b_1$. On the other hand,
$Q_2$ is associated with the tensor asymmetry in the transverse
plane. The $Q_1$ polarization is very peculiar. The deuteron
is polarized with the angle 45$^\circ$ or 135$^\circ$ with respect
to the longitudinal axis, and the cross section difference is taken.
This kind of peculiar asymmetry does not exist, of course, for the proton.
This ``intermediate" (according to Ref.\cite{hk}) polarization asymmetry
could lead to a new unique field of spin physics. A parton-model analysis
suggests that the asymmetry $Q_0$ should be related to $b_1$ type tensor  
distributions as \cite{hk}
\begin{equation}
A_{UQ_0}  =  \frac{\sum_a e_a^2 \, 
                  \left[ \, f_1^a (x_1) \, \bar b_1^a (x_2)
                          + \bar f_1^a (x_1) \, b_1^a (x_2) \, \right] }
                {\sum_a e_a^2 \, 
                  \left[ \, f_1^a (x_1) \, \bar f_1^a (x_2)
                          + \bar f_1^a (x_1) \, f_1^a (x_2) \, \right] }
                          \ ,
\label{eqn:auq0}
\end{equation}
where $f_1^a$ and $b_1^a$ are unpolarized and tensor-polarized
distributions for the quark $a$, and $\bar f_1^a$ and $\bar b_1^a$
are those for the antiquark. The momentum fractions are given by
$x_1$ and $x_2$ in the proton and the deuteron.
For example, $b_1^a$ is given by $b_1^a = [ q_a(0) -(q_a(+1)+q_a(-1))/2]/2$. 
We mentioned the sum rule $\int dx b_1 (x)=0$.
However, this relation crucially depends on the antiquark tensor
polarization $\bar b_1^a$, which cannot be well probed in the electron
scattering. One of the major advantages of using the $pd$ reaction is to
find $\bar b_1^a$ rather easily, at least according to the formalism.
The situation is similar to the case that the detailed 
$x$ dependence of $\bar u/\bar d$ was clarified
by the Fermilab-E866 Drell-Yan experiment although the difference
was originally indicated by the NMC data \cite{skpr}.
For example, in the large-$x_F$ region, Eq.(\ref{eqn:auq0}) becomes
\begin{equation}
A_{UQ_0} \textrm{(large $x_F$)} 
      \approx \frac{\sum_a e_a^2 \, f_1^a (x_1) \, \bar b_1^a (x_2)}
                   {\sum_a e_a^2 \, f_1^a (x_1) \, \bar f_1^a (x_2)}
\ .
\end{equation}
Therefore, the antiquark tensor distributions $\bar b_1^a$
can be determined if the unpolarized distributions are well
known in the proton and deuteron.

Another advantage of the $pd$ Drell-Yan is that the $\bar u/\bar d$
asymmetry can be investigated for the transversity distributions
\cite{km}. Because they have chiral-odd property, they cannot be
investigated in the inclusive electron scattering.
The $pd$/$pp$ Drell-Yan ratio is given by \cite{km}
\begin{align}
R_{pd} \equiv \frac{     \Delta_{(T)} \sigma_{pd}}
                   {2 \, \Delta_{(T)} \sigma_{pp}}
       & =     \frac{ \sum_a e_a^2 \, 
    \left[ \, \Delta_{(T)} q_a(x_1) \, 
              \Delta_{(T)} \bar q_a^{\, d}(x_2)
            + \Delta_{(T)} \bar q_a(x_1) \, 
              \Delta_{(T)} q_a^d(x_2) \, \right] }
              { 2 \, \sum_a e_a^2 \, 
    \left[ \, \Delta_{(T)} q_a(x_1) \, 
              \Delta_{(T)} \bar q_a(x_2)
            + \Delta_{(T)} \bar q_a(x_1) \, 
              \Delta_{(T)} q_a(x_2) \, \right] }
\nonumber \\
     & \rightarrow  \frac{1}{2} \, \left [ \, 1 
                 + \frac{\Delta_{(T)} \bar d (x_2)}
                        {\Delta_{(T)} \bar u (x_2)} 
                    \, \right ]_{x_2\rightarrow 0}
\ \ \ \text{for $x_F \rightarrow 1$}
\ ,
\end{align}
\begin{wrapfigure}{r}{0.46\textwidth}
   \vspace{-0.3cm}
   \begin{center}
       \epsfig{file=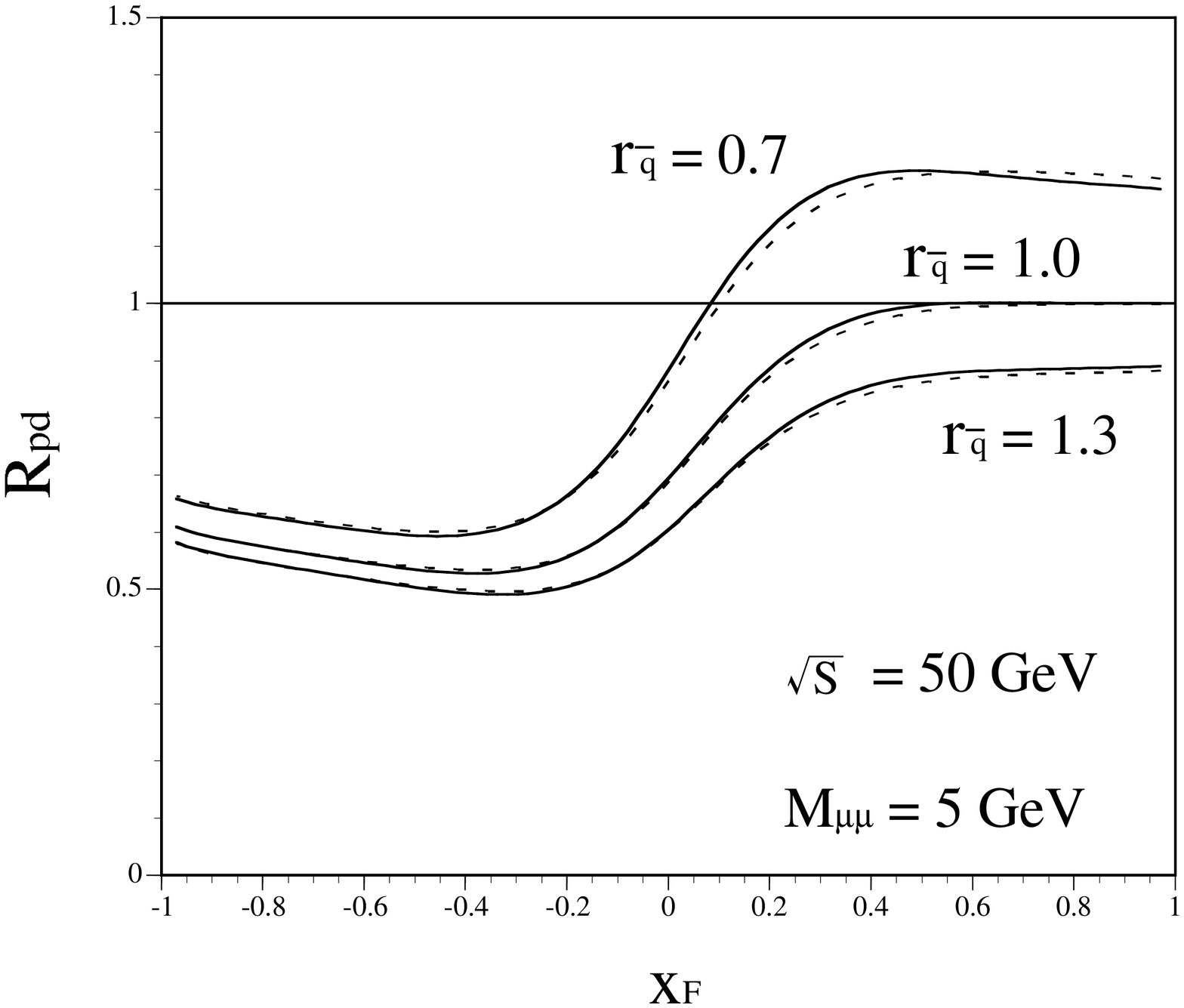,width=5.5cm} \\
       \vspace{-0.15cm}
       {\footnotesize {\bf FIGURE 2.} $pd/pp$ Drell-Yan ratio [5].}
   \end{center}
   \vspace{-0.5cm}
\end{wrapfigure}
\noindent
where $\Delta_{(T)}=\Delta$ or $\Delta_T$ depending on
the longitudinal or transverse polarization,
and $\Delta_{(T)} \sigma$ indicates the polarized cross-section difference.
In the second line, $x_F \rightarrow 1$ limit is taken together
with the assumption 
$\Delta_{(T)} u_v (x \rightarrow 1) \gg 
 \Delta_{(T)} d_v (x \rightarrow 1)$.
Leading-order numerical results are shown in Fig.2, where
$r_{\bar q} \equiv \Delta_{(T)} \bar u/ \Delta_{(T)} \bar d$.
It obviously indicates that the ratio is suitable for finding
the polarized light antiquark distributions, particularly
in the large-$x_F$ region.

In this way, we found that the $pd$ Drell-Yan is an alternative way
of studying the deuteron spin structure to the electron scattering.
Because other studies of spin-1 physics are not discussed 
in this paper, the author hopes that the interested reader will
look at the papers in the extensive lists of Ref.\cite{hk,km}.

\section*{Summary}

We explained the possibilities of investigating the polarized deuteron.
Exploring a new field of hadron spin, namely the tensor structure,
we should be able to test our knowledge of high-energy spin physics.
Because it has not been well investigated yet, we may encounter 
an unexpected result. Depending on the HERMES measurement in the
near future, this kind of topic could become an interesting area
of hadron physics. We also expressed the importance of polarized
proton-deuteron reactions, which are possible not only in collider
experiments but also in fixed-target ones.



\end{document}